\documentclass[11pt]{article}
\pdfoutput=1
\usepackage{graphicx}
\usepackage{epsfig}
\usepackage{amsfonts}
\usepackage{amscd}
\usepackage{latexsym}
\usepackage{amsmath,amssymb}
\usepackage{verbatim}
\usepackage{setspace}
\usepackage{subfigure}
\usepackage{color}
\usepackage{MnSymbol}
\usepackage{fancyhdr}
\usepackage{cite}
\usepackage{hyperref}
\usepackage{tensor}
\usepackage{xfrac}
\usepackage[textheight=9in, textwidth=6.5in, letterpaper]{geometry}
\usepackage{xcolor}
\usepackage{physics}
\usepackage{tcolorbox}
\def\p{\partial}
\usepackage{soul}
\usepackage{ulem}

\newcommand{\mS}{\mathcal{S}}

\newcommand{\erin}[1]{{\color{blue}[Erin: #1]}}

\def\ee{\end{equation}}
\def\be {\begin{equation}}
\def\ohh{\mathcal{O}^{\pm,h,\bar{h}}}
\def\co{\mathcal{O}}
\def\ch{\cal{H}}
\def\ci{\cal{I}}

\numberwithin{equation}{section}

{\cfoot{\thepage}}
\begin{document}
\begin{titlepage}
\unitlength = 1mm~\\
\vskip 3cm
\begin{center}
{\LARGE{\textsc{State-Operator Correspondence in Celestial Conformal Field Theory}}}

\vspace{0.8cm}
Erin Crawley{$^{*}$}, Noah Miller{$^{*}$}, Sruthi A. Narayanan{$^{*}$}, and Andrew Strominger{$^{*}$}\\
\vspace{1cm}

{$^*$\it  Center for the Fundamental Laws of Nature, Harvard University,\\
Cambridge, MA 02138, USA}

\vspace{0.8cm}

\begin{abstract}

The bulk-to-boundary dictionary for 4D celestial holography is given a new entry defining  2D boundary states living  on oriented circles on the celestial sphere. The states  are constructed using the 2D CFT state-operator correspondence from operator insertions corresponding to either incoming or outgoing  particles which cross the celestial sphere  inside the circle. The BPZ construction is applied to give an inner product on such states whose associated bulk adjoints are shown to involve a shadow transform. Scattering amplitudes are then given by BPZ inner products between states living on the same circle but with opposite orientations. 2D boundary states are found to encode the same information as their 4D bulk counterparts, but organized  in a radically different manner. 

\end{abstract}

\end{center}

\end{titlepage}

\tableofcontents

\section{Introduction}

The symmetries of every asymptotically flat 4D bulk quantum theory of gravity include the local conformal group acting in antipodal unison on the past and future celestial spheres at null infinity~\cite{Cachazo:2014fwa,Kapec:2014opa}. This implies that the $\cal S$-matrix can be holographically recast as boundary correlators of a 2D ``celestial conformal field theory" (CCFT) living on the celestial sphere~\cite{He:2015zea,Kapec:2016jld,Bagchi:2016bcd,Cheung:2016iub,Pasterski:2016qvg,Cardona:2017keg,Pasterski:2017kqt}.  The operators inserted in the correlator can correspond to both incoming and outgoing particles crossing the celestial sphere. The so-defined CCFT correlators enjoy some, but not all, properties of those in a garden-variety 2D CFT. It is an important open challenge  to find an intrinsic construction of any CCFT ({\it i.e.} other than as a transformation of a bulk theory)  starting with a microscopic theory such as string theory. However, it is implausible that such a construction will be possible for the real world any time soon as it would amount to a complete knowledge of all the laws of physics. Nevertheless, many properties of CCFT can be deduced from their rich symmetry properties and logical self-consistency conditions.

In this paper we deduce  entries in the celestial holographic  bulk-to-boundary dictionary concerning  the relation between the 4D bulk and 2D boundary Hilbert spaces, inner products and scattering problems. In the familiar case of AdS/CFT holography~\cite{Aharony:1999ti}, there is a very simple correspondence between bulk and boundary states: they are different descriptions of the same thing. Such a simple relation cannot possibly hold in celestial holography as the boundary is Euclidean while the bulk is Lorentzian.  A pair of 2D states can be defined by dividing the celestial sphere into northern and southern hemispheres. The northern and southern operator insertions - both  incoming and outgoing - then define ``northern" and ``southern" states on the celestial sphere. The inner product of a northern and southern 2D state is then a scattering amplitude.  Hence,  boundary states exist along with bulk states but they organize the information about the theory in a fascinating and very different manner. 
  
To construct a 2D inner product it is insightful to first understand the 4D bulk product from which it can be derived. The Klein-Gordon inner product provides a natural positive-definite norm on the bulk Hilbert space. However, it is unnatural from the boundary perspective both because it gives delta functions on the celestial sphere for conformal primaries~\cite{Pasterski:2016qvg,Pasterski:2017kqt} and because the conformal generators have non-standard adjoint properties~\cite{Bousso:2001mw,Ng:2012xp}.  A second conserved inner product is given by the classic Belavin, Polyakov and Zamolodchikov (BPZ) construction~\cite{Belavin:1984vu}, which starts with a CFT$_2$ two-point function  with  operators inserted at  opposite poles. This construction applies directly to CCFT in which the CFT$_2$ two-point function is given by the two-particle $\cal S$-matrix written in the conformal basis. The resulting inner product is indefinite, which is to be expected since states in the CCFT can have negative or complex conformal weights. We show that from the bulk perspective the BPZ adjoint of a 2D conformal primary state thereby involves a shadow transform~\cite{Ferrara:1972uq,SimmonsDuffin:2012uy} replacing  the standard Klein-Gordon complex conjugation. Shadowing one of the primary fields in the two-point function transforms the delta function to the more familiar CFT$_2$ power law. Shadows have been ubiquitous in discussions of CCFT,\footnote{An important related question, not addressed herein, is what criteria determine an optimal ``complete"  set of operators. Whereas independently the primaries and shadowed primaries provide complete bases of square normalizable radiative wave functions, they each appear naturally in different contexts. At the same time,  the primaries and their shadows with arbitrary conformal weights clearly comprise an over-complete set. We hope the observations of this paper prove useful in addressing this question.} indeed in~\cite{Fan:2021isc} the scattering of shadow states was recently derived and found to have elegant factorization properties.    
    
This paper is organized as follows. Section~\ref{sec:bulkinner} begins with a review of conserved 4D bulk inner products, including the origin of delta functions in the Klein-Gordon inner product of conformal primaries.  The conserved bulk shadow  product, in which the shadow replaces complex conjugation, is introduced and computed for conformal primaries. It is shown to have both the familiar boundary power-law behavior dictated by conformal invariance and the familiar  adjoint relations for conformal generators. In section~\ref{sec:boundary} we describe how conformal invariance allows us to associate a state living on the circle surrounding every operator insertion. These states are defined explicitly using the operators appearing in the mode expansions of conformal primaries. We then construct the BPZ  inner product of these states from the two-point function, and show that for CCFT the adjoint state involves a shadow transform.  Viewed as a bulk inner product on single-particle states, we recover the shadow product of section~\ref{sec:bulkinner}.  We close in section~\ref{sec:bulkvboundary} with a discussion of  the holographic relation, in the multi-particle context, between the 2D boundary Hilbert space and the 4D bulk Hilbert space. We also relate  the boundary scattering problem, which concerns a map between ``northern" and ``southern" states on the respective hemispheres of the celestial sphere, with the bulk scattering problem which concerns a map from past to future null infinity. 
  
Throughout this paper we use the terms ``inner product" and ``Hilbert space" to include indefinite inner products and Hilbert spaces.

\section{Bulk  inner products}\label{sec:bulkinner}
In order to define an inner product on the 2D boundary, we must first discuss the 4D bulk products to which they are related. In this section we discuss two standard products on the 4D bulk space, the Klein-Gordon and symplectic, as well as a modification of them involving the shadow transform which will prove to be useful in the subsequent discussion about 2D CCFT inner products.

\subsection{Symplectic and Klein-Gordon products}  
Consider 4D Minkowski space with the standard metric 
\begin{equation}
ds^2 = \eta_{\mu\nu}dX^\mu dX^\nu = -\left(dX^0\right)^2+\left(dX^1\right)^2+\left(dX^2\right)^2+\left(dX^3\right)^2 ,
\end{equation}
and let $\Phi_{1,2}$ denote any two solutions of the scalar wave equation. From these one can construct the symplectic current
\be J_\mu(\Phi_1,\Phi_2)=\Phi_1\overleftrightarrow \p_\mu \Phi_2 \equiv \Phi_1(\partial_\mu\Phi_2)-(\partial_\mu\Phi_1)\Phi_2,
\ee
which is conserved: 
\be 
\p^\mu J_\mu=0.
\ee
The symplectic product of two wavefunctions is accordingly defined as 
\begin{equation}
(\Phi_1,\Phi_2)_{\rm sym} = \int_{\Sigma_3} d^3\Sigma^\mu J_\mu(\Phi_1,\Phi_2).
\end{equation}
This product is conserved in the sense that it is independent of the choice of complete spacelike slice $\Sigma_3$.  
Since $J_\mu$ is conserved for any pair of solutions, there are many possible conserved scalar products, a freedom which is exploited below. For example, the usual  Klein-Gordon product is obtained from the symplectic product as
\begin{equation}
(\Phi_1,\Phi_2)_{\rm{KG}} =-i\int_{\Sigma_3} d^3\Sigma^\mu J_\mu(\Phi_1,\Phi_2^*) =-i(\Phi_1, \Phi_2^\ast)_{\rm{sym}},
\end{equation}
and is also conserved. One can use the Klein-Gordon product to construct a positive definite 4D Hilbert space in the usual way. Similar inner products can be constructed from symplectic structures~\cite{Ashtekar:1987tt} for arbitrary integer spin $J$. The treatment of half-integer spin $J$ requires an inner product like the Dirac inner product considered in~\cite{Fotopoulos:2020bqj,Narayanan:2020amh,Mueck:2020wtx}.

\subsection{Lorentz generators and their adjoints}
We will be interested in the action of the Lorentz group on these wavefunctions which is infinitesimally generated by the Lie action of the six vector fields
\begin{eqnarray}
\xi_1 & = & -\frac{1}{2}\left[(X^1+iX^2)(\partial_0+\partial_3) +(X^0-X^3)(\partial_1+i\partial_2)\right] \cr
\xi_0  & = & -\frac{1}{2}\left[X^3\partial_0 - i X^2\partial_1+iX^1\partial_2+X^0\partial_3\right]\cr
\xi_{-1}  & = &\frac{1}{2}\left[(X^1-iX^2)(\partial_0-\partial_3) + (X^0+X^3)(\partial_1-i\partial_2)\right]\cr
\bar\xi_1 & = & -\frac{1}{2}\left[(X^1-iX^2)(\partial_0+\partial_3) +(X^0-X^3)(\partial_1-i\partial_2)\right] \cr
\bar\xi_0  & = & -\frac{1}{2}\left[X^3\partial_0 + i X^2\partial_1-iX^1\partial_2+X^0\partial_3\right]\cr
\bar\xi_{-1}  & = &\frac{1}{2}\left[(X^1+iX^2)(\partial_0-\partial_3) + (X^0+X^3)(\partial_1+i\partial_2)\right] 
\end{eqnarray}
obeying $\bar{\xi}_n = (\xi_n)^\ast$. Denoting their Lie action by  $ L_n = -\mathcal{L}_{\xi_n}, \bar{L}_n = -\bar{\mathcal{L}}_{\bar{\xi}_n}, $ one finds 
\begin{equation}
[ L_m, L_n]=(m-n) L_{m+n},~~~~ [\bar{L}_m,\bar{L}_n]=(m-n)\bar{L}_{m+n}.
\end{equation}
It is straightforward to verify that~\cite{Bousso:2001mw}
\begin{equation}
(\Phi_1,L_n\Phi_2)_{\rm{KG}} = -(\bar{L}_n\Phi_1,\Phi_2)_{\rm{KG}}
\end{equation}
for $\Phi_{1,2}$, which decay sufficiently rapidly at infinity. Hence with respect to this scalar product the adjoint is 
\begin{equation}
\label{adj} L_n^\dagger = -\bar{L}_n.
\ee
Note that this is not the standard adjoint relation $L_n^\dagger = L_{-n}$ which one would expect in CFT$_2$. However, when we define our celestial inner product, we will indeed recover the more standard adjoint equation in that setting.

\subsection{Conformal primary wavefunctions}
Bulk 4D $\mathcal{S}$-matrix elements in momentum space can be recast as ``celestial amplitudes" that transform like 2D conformal correlators by way of a Mellin transform.\footnote{In the remainder of this work we focus on massless wavefunctions but our results can be generalized to the massive case as well. In the massive case the transformation is an integral over a hyperbolic slice rather than a Mellin transform.} Celestial amplitudes are constructed by scattering states which are SL$(2,\mathbb{C})$ primary wavefunctions rather than momentum space plane waves~\cite{Pasterski:2016qvg,Pasterski:2017kqt}. These states $\phi_{\pm,\Delta, J}^{\mu_1\ldots \mu_s}\left(X^\mu;w,\bar{w} \right)$, referred to as conformal primary wavefunctions, are labelled by an arbitrary complex conformal  dimension $\Delta$,  spin $J$, ingoing/outgoing subscript  $\pm$ and a point $(w,\bar w)$ where they cross the celestial sphere. In what follows we label them by their conformal weights $(h, \bar{h})$, related in the usual way by 
\be 
(h,\bar h)=\left(\frac{1}{2}(\Delta+J),\frac{1}{2} (\Delta-J)\right).
\ee 
Under Lorentz and conformal transformations primary wavefunctions satisfy 
\begin{equation}\label{pc}
\Lambda_{\nu_1}^{\mu_1}\ldots\Lambda_{\nu_s}^{\mu_s} \phi_{\pm,h,\bar h}^{\nu_1\ldots \nu_s}\left((\Lambda^{-1})^\mu_\nu X^\nu; w, \bar{w} \right) = (cw+d)^{-2h}(\bar{c}\bar{w}+\bar{d})^{-2\bar{h}}\phi_{\pm,h,\bar h}^{\mu_1\ldots \mu_s}\left(X^\mu; \frac{aw+b}{cw+d},\frac{\bar{a}\bar{w}+\bar{b}}{\bar{c}\bar{w}+\bar{d}}  \right).
\end{equation} 
The infinitesimal version of this equation is
\be\label{ptr} 
L_n\phi_{\pm ,h,\bar h}^{\mu_1\ldots\mu_s}(X^\mu,w,\bar w)=(w^{n+1}\partial_w+h(n+1)w^n)\phi^{\mu_1\ldots \mu_s}_{\pm,h,\bar h}(X^\mu,w,\bar w),
\ee 
and likewise for $\bar{L}_n$. The continuum of modes obtained by Mellin transforms on the unitary principal series $\Delta=1+i\lambda$ for\footnote{For massive wavefunctions this is restricted to $\lambda\in\mathbb{R}_{\geq 0}$.} $\lambda\in\mathbb{R}$ are of special interest~\cite{Pasterski:2017kqt} because they comprise a complete basis of solutions for normalizable radiative wave packets.\footnote{They do not however form representations of the Poincar\'e group. Another basis of interest~\cite{Atanasov:2021oyu} which do furnish Poincar\'e representations, and moreover include the soft currents~\cite{Guevara:2021abz} are those with real integral $\Delta$. In this paper we assume that $\Delta$ lies on the unitary principal series but many of our formulae can be generalized to the integral case.} The symplectic product between two integer spin conformal primary wavefunctions on the unitary principal series is
\begin{equation}\label{symplectic}
\left(\phi_{\pm,h_1,\bar h_1}^{\mu_1\ldots \mu_s}(X^\mu;w_1,\bar{w}_1),\phi_{\mp,h_2,\bar h_2}^{\nu_1\ldots\nu_s}(X^\mu;w_2,\bar{w}_2)\right)_{\rm sym}=\mathcal{C}^\pm_{J_1}(\lambda_1)\delta_{J_1-J_2}\delta(\lambda_1+\lambda_2)\delta^{(2)}(w_1-w_2) ,
\end{equation}
where $\mathcal{C}^\pm_{J_1}(\lambda_1)$ is a normalization that can be found in~\cite{Pasterski:2017kqt}.

As an explicit example, for the case of massless scalars,  conformal primary wavefunctions take the form 
\be 
\varphi_{\pm, h,\bar{h}}(X^\mu;\vec{w}) = \frac{(\mp i)^\Delta\Gamma(\Delta)}{(-q(\vec{w})\cdot X\mp i\epsilon)^\Delta},
\ee
where $q(\vec{w}) = (1+w\bar{w},w+\bar{w},i(\bar{w}-w), 1-w\bar{w})$ is a null vector which points to $w=\frac{q^1+iq^2}{q^0+q^3}$ on the celestial sphere. The symplectic product is 
\be\label{scl}
\left(\varphi_{\pm,h_1,\bar{h}_1}(X^\mu;w_1,\bar{w}_1),\varphi_{\mp,h_2,\bar{h}_2}(X^\mu;w_2,\bar{w}_2)\right)_{\rm sym} = \pm 8i\pi^4\delta(\lambda_1+\lambda_2)\delta^{(2)}(w_1-w_2),
\ee
so we see that the normalization is $\mathcal{C}^\pm_0(\lambda_1) = \pm 8i\pi^4$. Likewise the Klein-Gordon product is 
\be\label{kgl}
(\varphi_{\pm,h_1,\bar{h}_1}(X^\mu;w_1,\bar{w}_1),\varphi_{\pm,h_2,\bar{h}_2}(X^\mu;w_2,\bar{w}_2))_{\rm sym} = \pm8\pi^4\delta(\lambda_1-\lambda_2)\delta^{(2)}(w_1-w_2).
\ee

\subsection{Shadow product}\label{sec:shadow}

One of the goals of rewriting 4D scattering amplitudes in a conformal basis is to apply the powerful techniques of CFT$_2$ to analyze and constrain them. For a free theory, as we consider in this work, the momentum-space $\mathcal{S}$-matrix elements for one incoming and one outgoing particle are given by  the Klein-Gordon product of  plane waves. The Mellin transform of these $\mathcal{S}$-matrix elements are bulk products of conformal primary wavefunctions and can be identified with a CFT$_2$ two-point function. However, while the result~\eqref{symplectic} is certainly fully conformally invariant, one expects to see a factor $(z-w)^{-2h}(\bar z -\bar w)^{-2 \bar h}$ in a CFT$_2$ two-point function rather than the delta function $\delta^{(2)}(z-w)$.  Moreover,  the adjoint relation $L_n^\dagger=-\bar{L}_n$ derived from the Klein-Gordon inner product is not the one usually encountered in CFT$_2$. Rather we expect the conformal generators to obey $L_n^\dagger = L_{-n}$. 

Since both of these unfamiliar properties derive from the choice of inner product,  we here consider modified  inner products. If $\mathcal{M}$ is any map on the space of solutions, a new conserved product can be constructed from $J_\mu(\mathcal{M}\Phi_1,\Phi_2)$. Such modified products have been considered in many contexts, including holographic ones similar to the current case in~\cite{Witten:2001kn,Bousso:2001mw,Ng:2012xp,Jafferis:2013qia}. A useful choice of $\mathcal{M}$ is provided by the 2D shadow transform $\mathcal{S}$~\cite{Ferrara:1972uq,SimmonsDuffin:2012uy}, which takes $J\to-J$ and $\Delta\to 2-\Delta$ thereby taking a field of conformal weight $(h,\bar{h})$ to one of weight $(1-h,1-\bar{h})$. The transform also obeys $\mathcal{S}^2=(-1)^{2J}$ when acting on a conformal primary. The shadow transform, denoted by $\widetilde \phi$,  for a field of arbitrary spin is given\footnote{Using Gamma-function identities this definition is symmetric under $w\leftrightarrow \bar w, ~h\leftrightarrow \bar h$ up to a factor of $(-1)^{J}$.} by~\cite{Osborn:2012vt} 
\begin{equation}
\widetilde\phi_{\pm,1-h,1-\bar h}^{\mu_1\ldots\mu_s}(X^\mu; w,\bar{w}) = \frac{\Gamma(2-2\bar{h})}{\pi\Gamma(2h-1)}\int d^2z\frac{1}{(w-z)^{2-2h}(\bar{w}-\bar{z})^{2-2\bar{h}}}\phi_{\pm,h,\bar h }^{\mu_1\ldots\mu_s}(X^\mu;z,\bar{z}).
\end{equation}
Note that this acts only on the 2D celestial coordinates $(z, \bar z)$ and not the 4D bulk coordinates $X^\mu$. It also preserves the distinction between incoming and outgoing modes. $\phi$ and $\widetilde \phi$ provide  two distinct but equally complete bases for normalizable massless scalar wave packets~\cite{Pasterski:2017kqt}. The first complete basis, as exhibited in~\eqref{scl}, are simple Mellin transforms of plane waves while the second complete basis consists of the shadow transformations thereof. The shadow transform of any normalizable solution of the wave equation is defined by first decomposing it into a basis of $\phi$ conformal primaries and then shadowing the individual components.

We define the conserved shadow product between any two modes by 
\begin{equation}\label{shd}
(\Phi_1,\Phi_2)_{\mS} =\int_{\Sigma_3} d^3\Sigma^\mu J_\mu( \widetilde\Phi_1,\Phi_2) =(\widetilde\Phi_1, \Phi_2)_{\rm{sym}}=i (\widetilde\Phi_1, \Phi_2^*)_{\rm{KG}}.
\end{equation}
For conformal primaries this becomes \begin{equation}\label{eq:shadowsym}
\left( \phi_{\pm,h_1,\bar h_1}^{\mu_1\ldots\mu_s}(X^\mu;w_1,\bar{w}_1),  \phi^{\nu_1\ldots \nu_s}_{\mp,h_2,\bar h_2}(X^\mu;w_2,\bar{w}_2) \right)_{\mathcal{S}}=\frac{\Gamma(2-2\bar{h}_1)}{\pi\Gamma(2h_1-1)}\frac{\delta_{h_1-\bar h_1+h_2-\bar h_2}\mathcal{C}^\pm_{J_1}(\lambda_1)\delta(\lambda_1+\lambda_2)}{(w_1-w_2)^{2h_2}(\bar{w}_1-\bar{w}_2)^{2\bar{h}_2}}.
\end{equation}
Note that because of the action of $\mathcal S$ this enforces $h_1 = 1 - h_2$ (or $\lambda_1=-\lambda_2$) instead of the usual $h_1 = h_2$.

The shadow product \eqref{eq:shadowsym} is non-vanishing at non-coincident points and has the power law behavior that is more familiar in CFT$_2$. 

\section{Boundary}\label{sec:boundary}

In AdS/CFT holography, there is a very simple correspondence between bulk and boundary states: boundary fields are identified with the boundary value of solutions to the bulk equations of motion. An important ingredient for this correspondence is that both spaces are Euclidean. However, in celestial holography the boundary is Euclidean while the bulk is Lorentzian, which suggests that a similarly simple relation cannot possibly hold. Nonetheless, there is still a correspondence between bulk and boundary states, albeit less direct since the bulk and boundary descriptions organize the information about the theory in different ways. In this section we use the 2D state-operator correspondence to describe boundary states in CCFT. We then use the BPZ construction to define an inner product between these states and show that the result corresponds to the bulk shadow product.  This gives an indefinite norm on 2D states with respect to which $L_n^\dagger = L_{-n}$.

\subsection{State-operator correspondence in CCFT}

Boundary operators for massless fields in CCFT are defined as Mellin transforms (or shadows thereof) of 4D bulk momentum eigenoperators, and their correlators are defined by Mellin transforms  of the corresponding  momentum-space  scattering amplitudes. In this subsection we describe the 2D boundary states related to these boundary operators following the usual CFT$_2$ state-operator correspondence.

Let $\co^{+,h,\bar h}(w, \bar w)$ ($\co^{-,h,\bar h}(w, \bar w)$)
denote an $X^\mu$-independent operator of weight $(h,\bar h)$ constructed via the Klein-Gordon product of a positive (negative) conformal primary wavefunction with the quantum field operator $\hat \Phi$,
\be \co^{\pm, h,\bar h}(w, \bar w)=\left( \phi^{\mu_1 \ldots \mu_s}_{\pm,h,\bar h}(w,\bar w), \hat \Phi^{\nu_1 \ldots \nu_s}\right)_{\rm KG}.
\ee
Similar constructions have been done explicitly in~\cite{Donnay:2018neh} where for arbitrary spin one must consider the appropriate inner product, \textit{i.e.} the Maxwell inner product for spin-1. These are creation (annihilation) operators in the 4D Hilbert space for incoming (outgoing) particles which cross the celestial sphere at the point $(w, \bar w)$. The 2D state-operator correspondence gives  an associated  2D Hilbert space of states which live on the small circle surrounding $(w, \bar w)$.  In the boundary picture $\co^{+,h,\bar h}$and $\co^{-,h,\bar h}$ create distinct states on this circle.  This is different  from the bulk picture in which $\co^{-,h,\bar h}$ annihilates the state created by $\co^{+,h,\bar h}$.  So, unlike in traditional AdS/CFT, bulk and boundary states are not in exact correspondence although they are derived from the same set of operators.

We denote the 2D state associated to $\co^{\pm, h,\bar h}$ as  $\mathcal{O}^{\pm, h,\bar{h}}(w,\bar{w})|0_2\rangle$, where $|0_2\rangle$ denotes the 2D vacuum.\footnote{We remind the reader that $\ket{0_2}$ is not a state living in the 4D bulk Hilbert space, but in the distinct 2D boundary Hilbert space.} It is useful to perform a mode expansion of such states around one pole $(w,\bar w)=0$ of the celestial sphere:
\begin{equation}\label{inmodeexpansion}
 \mathcal{O}^{\pm, h,\bar{h}}(w,\bar{w})|0_2\rangle  = \sum_{m,\bar{m}} \frac{\mathcal{O}_{m,\bar{m}}^{\pm, h,\bar{h}}}{w^{h+m}\bar{w}^{\bar{h}+\bar{m}}}|0_2\rangle, \ \ (m,\bar{m})\in\mathbb{Z}-(h,\bar{h}).
\end{equation}
where the constraints on the allowed values of $(m,\bar m)$ follow from the demand that the state has a well-defined Laurent expansion about $(w,\bar{w})=0$. Formally treating $w$ and $\bar w$ as independent complex variables, the modes  are given by contour integrals
\begin{equation}\label{inmodes}
\mathcal{O}_{m,\bar{m}}^{\pm, h,\bar{h}}|0_2\rangle = \frac{1}{(2\pi i)^2}\oint_0 dw\oint_0 d\bar{w} w^{h+m-1}\bar{w}^{\bar{h}+\bar{m}-1}\mathcal{O}^{\pm, h,\bar{h}}(w,\bar{w})|0_2\rangle,
\end{equation}
where the contour is a small circle around $(w,\bar{w})=0$. The $\mathcal{O}_{m,\bar{m}}^{\pm, h,\bar{h}}$ modes are operators that act on 2D states on the circle surrounding the location of the operator insertion. The primary state in the 2D CCFT associated to  $\ohh$ is then defined as
\begin{equation}\label{inket}
|h,\bar{h},\pm\rangle = \lim_{w,\bar{w}\rightarrow 0}\mathcal{O}^{\pm, h,\bar{h}}(w,\bar{w})|0_2\rangle = \mathcal{O}_{-h,-\bar{h}}^{\pm, h,\bar{h}}|0_2\rangle,
\end{equation}
where the presumption that the expansion~\eqref{inmodeexpansion} is finite in this limit implies that 
\begin{equation}\label{eq_cftannihilation}
\mathcal{O}_{m,\bar{m}}^{\pm, h,\bar{h}}|0_2\rangle =0, \ \mbox{for} \ m+h,\bar{m}+\bar{h}>0. 
\end{equation}
Hence the above range of modes serve as annihilation operators acting on $|0_2\rangle$. 

In typical CFT$_2$ fashion one can act on this primary state with $L_n$ to obtain the descendants $\mathcal{O}_{n-h,-\bar{h}}|0_2\rangle$. The action of $L_n$ on primary fields can be derived from the OPE of the 2D boundary stress tensor with $\ohh$
which is~\cite{Kapec:2016jld,Fotopoulos:2019} 
\be T_{zz}(z)\mathcal{O}^{\pm, h,\bar{h}}(w,\bar{w})= \frac{h}{(z-w)^2}\mathcal{O}^{\pm, h,\bar{h}}(w,\bar{w})+ \frac{1}{z-w}\partial_w\mathcal{O}^{\pm, h,\bar{h}}(w,\bar{w})+\ldots 
\ee
where the ellipsis denotes regular terms. One then has~\cite{DiFrancesco:639405}
\begin{eqnarray} 
[L_n,\mathcal{O}^{\pm, h,\bar{h}}(w,\bar{w})] & = & \oint_w dz \frac{z^{n+1}}{ 2\pi i}T_{zz}(z)\mathcal{O}^{\pm, h,\bar{h}}(w,\bar{w})\cr &=&h(n+1)w^n\mathcal{O}^{\pm, h,\bar{h}}(w,\bar{w})+w^{n+1}\partial_w\mathcal{O}^{\pm, h,\bar{h}}(w,\bar{w}). 
\end{eqnarray}
This transformation law of course agrees with that derived from the transformation law for primary wavefunctions in \eqref{ptr}. 
Using the mode expansion and SL$(2,\mathbb{C})$ invariance of the vacuum $L_n |0_2\rangle =0$ it follows that    
\be  \label{Lnmodes}
L_n\mathcal{O}_{m,\bar{m}}^{\pm, h,\bar{h}}|0_2\rangle = ((h-1)n-m)\mathcal{O}_{m+n,\bar{m}}^{\pm, h,\bar{h}}|0_2\rangle, 
\ee
with similar relations for $\bar L_n$. Hence by action with powers of $L_{-1}=\p_w,~~ \bar L_{-1}= \p_{\bar w}$ we can, as usual, construct all descendants of the primary state $|h,\bar{h},\pm\rangle$. Within some radius of convergence around the pole  these provide a complete basis of single-particle states associated with the operator $\ohh$. It follows that 2D states  of the general form
\begin{equation}\label{rdl}
\mathcal{O}_{m_1,\bar{m}_1}^{\pm, h_1,\bar{h}_1}\ldots\mathcal{O}_{m_N,\bar{m}_N}^{\pm, h_N,\bar{h}_N}|0_2\rangle, \ \mbox{for} \ m_k+h_k,\bar{m}_k+\bar{h}_k\le 0. 
\end{equation}
similarly form a complete basis for $N$-particle states within some radius of convergence. Beyond the free field theory case considered here, the operators will have non-trivial OPEs which must be taken into account in describing such states. 
\begin{figure}[htb]
\begin{center}
\includegraphics[scale=0.5]{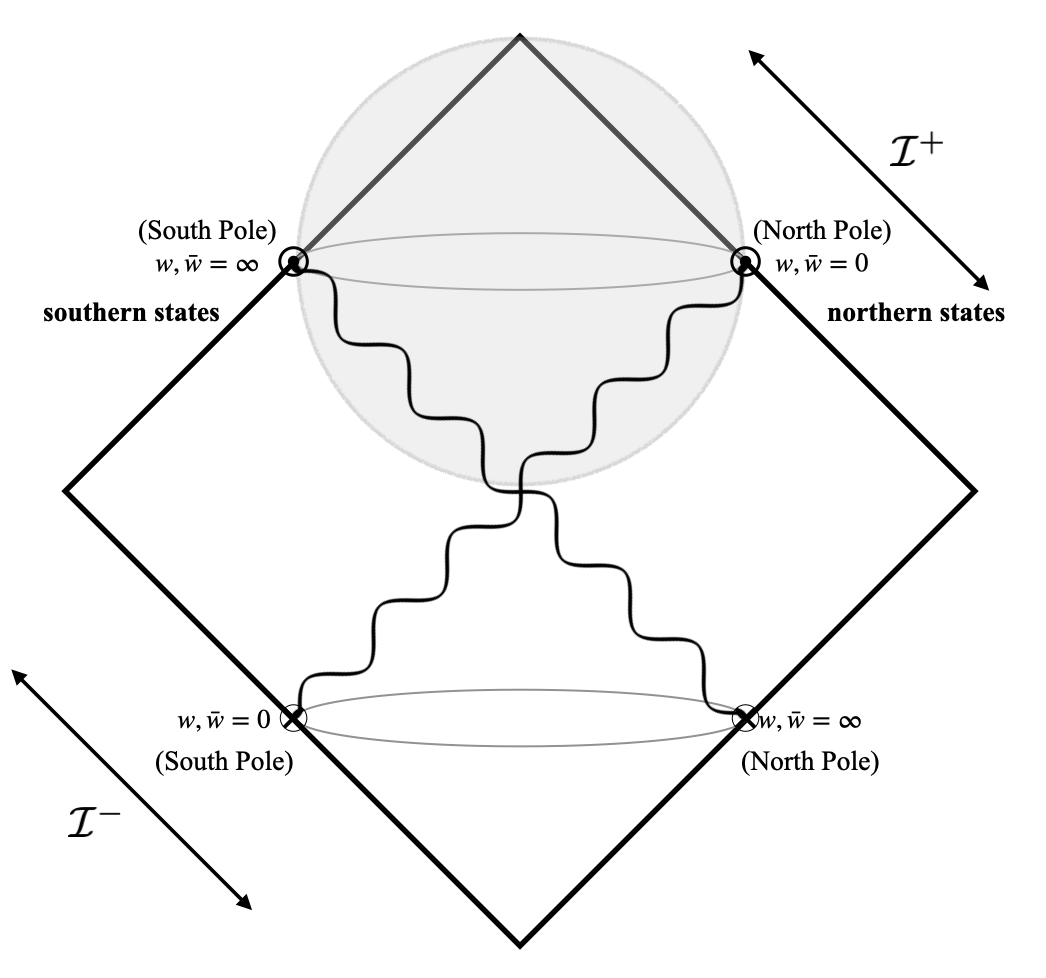}
\end{center}
\caption{``Northern" states enter in  the southern hemisphere on $\mathcal{I}^-$ and exit in the northern one on  $\mathcal{I}^+$. ``Southern" states enter in  the northern hemisphere on $\mathcal{I}^-$ and exit in the southern one on  $\mathcal{I}^+$. Due to the antipodal identification between the angular coordinates at $\mathcal{I}^+$ and $\mathcal{I}^-$, a free massless particle will enter and exit at the same  point on the celestial sphere.} 
\label{northsouth}
\end{figure}

It will be useful to have a pictorial understanding of these states in relation to incoming and outgoing bulk particles. Massless particles enter at some point on $\mathcal{I}^-$ and exit at some point on $\mathcal{I}^+$. If we refer to $(w,\bar w)=0$ as the south pole on ${\cal I}^-$ and, due to the antipodal matching, the north pole on ${\cal I}^+$, these 2D states describe particles which enter near the south pole on  ${\cal I}^-$ and exit near the north pole on  ${\cal I}^+$. All such excitations are north-moving and we will accordingly refer to these  as ``northern states" as shown Figure~\ref{northsouth}. Such a description cannot be used for states corresponding to south-moving particles entering (exiting) on the north (south) pole  because they are  at $(w,\bar{w})=\infty$ as shown in Figure~\ref{northsouth}. As we will see in the next section, such ``southern states"  are naturally viewed as ``2D bra-states" rather than ``2D ket-states".  We avoid using the term in- or out-states here in order to avoid confusion with  the very different bulk notion of incoming or outgoing particles. 

\subsection{BPZ inner  product in  CCFT}

In this section we show how the classic construction of BPZ~\cite{Belavin:1984vu} defines  an indefinite 2D inner product for CCFT. Unfamiliar features arise since the fields have complex conformal weights. An operator acting on the bra-vacuum state $\langle 0_2|$ is defined starting with  
\begin{equation} \label{eq_outoperator}
    \mathcal{B}'^{\pm, h, \bar{h}}(z, \bar z) \equiv z^{2 h} \bar{z}^{2 \bar{h}} \mathcal{B}^{\pm, h, \bar{h}} (z, \bar{z}).
\end{equation}
where  $\mathcal{B}^{\pm, h,\bar{h}}$ is an arbitrary operator, notated as such to emphasize that it is distinct from the previously introduced $\mathcal{O}^{\pm,h,\bar{h}}$ and is associated with the bra-states.  Defining $z'=-\frac{1}{z}$, we see that $\left({\p z \over \p z'}\right)^h \left({\p \bar z \over \p \bar z'}\right)^{\bar h}\mathcal{B}^{\pm, h, \bar{h}}(z, \bar z)$ is just the conformally transformed operator in $z'$ coordinates  which should be smooth for finite $z'$. Hence the action of $\mathcal{B}'$ must be finite when acting on $\langle 0_2|$ as $\frac{1}{z}\rightarrow 0$. This requires that the bra-states admit a mode expansion around $z=\infty$ in integer powers of $(z,\bar{z})$ as
\begin{equation}\label{outmodeexpansion}
 \langle 0_2|\mathcal{B}'^{\pm, h,\bar{h}}(z,\bar{z}) = \langle 0_2|\sum_{n,\bar{n}}\frac{\mathcal{B}'^{\pm, h,\bar{h}}_{n,\bar{n}}}{ z^{n-h}\bar{z}^{\bar{n}-\bar{h}}}, \ \ (n,\bar{n})\in\mathbb{Z}+(h,\bar{h}).
\end{equation}
where we note the range of modes $(n, \bar{n})$ differs from the ket-state expansion~\eqref{inmodeexpansion}. Again the modes are given by formal contour integrals
\begin{equation}\label{outmodes}
\langle 0_2|\mathcal{B}'^{\pm, h,\bar{h}}_{n,\bar{n}} = \frac{1}{(2\pi i)^2}\oint_\infty dz\oint_\infty d\bar{z} z^{n-h-1}\bar{z}^{\bar{n}-\bar{h}-1}\langle 0_2|{\mathcal{B}'}^{\pm, h,\bar{h}}(z,\bar{z}) ,
\end{equation}
where the contour integrals are taken around $z,\bar{z}=\infty$. Finally, a primary bra-state is defined as 
\begin{equation}\label{outbra}
\langle h,\bar{h},\pm| = \langle 0_2|\lim_{z,\bar{z}\rightarrow \infty}\mathcal{B}'^{\pm, h,\bar{h}}(z,\bar{z}) = \langle 0_2|\mathcal{B}'^{\pm, h,\bar{h}}_{h,\bar{h}}.
\end{equation}
For the state~\eqref{outmodeexpansion} to not diverge in this limit, we must have 
\begin{equation}
\langle 0_2|\mathcal{B}'^{\pm, h,\bar{h}}_{n,\bar{n}}=0, \ \mbox{for} \ (n-h,\bar{n}-\bar{h})<0, 
\end{equation}
\textit{i.e.} these serve as annihilation operators on $\langle 0_2|$. 

In what follows, we will construct the inner product $\langle 0_2|{\mathcal{B}}^{\prime \pm h_1,\bar{h}_1}_{n,\bar{n}}\mathcal{O}_{m,\bar{m}}^{\mp, h_2,\bar{h}_2}|0_2\rangle $ by using the two-point function between $\mathcal{B}$ and $\mathcal{O}$ whose form, assuming $w_1\neq w_2$, is dictated by conformal invariance to be 
\begin{equation}\label{cft2point}
\langle 0_2| \mathcal{B}^{\pm, h_1,\bar{h}_1}(w_1,\bar{w}_1)\mathcal{O}^{\mp, h_2,\bar{h}_2}(w_2,\bar{w}_2)|0_2\rangle = \frac{\mathcal{C}\delta_{h_1,h_2}\delta_{\bar{h}_1,\bar{h}_2}}{(w_1-w_2)^{2h_1}(\bar{w}_1-\bar{w}_2)^{2\bar{h}_1}}.
\end{equation}
The constant $\mathcal{C}$ depends on the choice of operators but not on $w_1$ or $w_2$, and the expectation value is defined as the two-point scattering amplitude of the particles associated to the operators $\mathcal{B}$ and $\mathcal{O}$. For free field theory, as considered here, the 4D bulk momentum space scattering amplitude is the symplectic product of the corresponding plane waves up to a factor of $i$. Since Mellin transforms of 4D bulk momentum space scattering amplitudes are CCFT correlators, this two-point function equals, up to a factor of $i$, the symplectic product of the associated wavefunctions. Note that if we take $\mathcal{B}^{\pm,h,\bar{h}} = \mathcal{O}^{\mp,h,\bar{h}}$ the constant $\mathcal{C}$ vanishes, so the family of operators $\ohh$ cannot be  self-adjoint with respect to this inner product. In order for this to be nonzero we choose to use the shadow and take  
\begin{equation}\label{bo} 
\mathcal{B}^{\pm, h,\bar{h}}=\widetilde{ \mathcal{O}}^{\pm, 1-h,1-\bar{h}}.
\end{equation}
One then has 
\begin{align} \label{Stwopoint}
\langle 0_2|  \widetilde{\mathcal{O}}^{\pm, 1-h_1,1-\bar{h}_1}(w_1,\bar{w}_1)\mathcal{O}^{\mp, h_2,\bar{h}_2}(w_2,\bar{w}_2)|0_2 \rangle 
= & \left( \phi_{\pm,h_1,\bar h_1}^{\mu_1\ldots\mu_s}(X^\mu;w_1,\bar{w}_1),  \phi_{\mp,h_2,\bar h_2}^{\nu_1\ldots\nu_s}(X^\mu;w_2,\bar{w}_2) \right)_{\mS}\cr
= & \frac{\Gamma(2-2\bar{h}_1)}{\pi\Gamma(2h_1-1)}\frac{\delta_{h_1+-\bar h_1+h_2-\bar h_2}\mathcal{C}_{J_1}^\pm(\lambda_1)\delta(\lambda_1+\lambda_2)}{(w_1-w_2)^{2h_2}(\bar{w}_1-\bar{w}_2)^{2\bar{h}_2}}.
\end{align}

So far we only have a formula for correlation functions of primary operators and their primary shadows at non-coincident points on the sphere. We seek an inner product on primary states and their descendants where the latter follows from the former in the usual manner~\cite{Belavin:1984vu}. If we insert modes of $\widetilde\co'$ and $\co$ between vacuum states, we find
\begin{align}\label{innermodes}
\langle 0_2|{\widetilde{\mathcal{O}}}^{\prime \pm ,1-h_1,1-\bar{h}_1}_{n,\bar{n}}\mathcal{O}_{m,\bar{m}}^{\mp, h_2,\bar{h}_2}|0_2\rangle  =&  \frac{1}{(2\pi i)^4}\oint_\infty dw_1\oint_\infty d\bar w_1 \nonumber w_1^{n+h_1-2}\bar{w}_1^{\bar{n}+\bar{h}_1-2}\oint_0 dw_2 \oint_0 d\bar{w}_2 w_2^{h_2+m-1}\bar{w}_2^{\bar{h}_2+\bar{m}-1} \\
&\times w_1^{2(1-h_1)}\bar{w}_1^{2(1-\bar{h}_1)} \langle 0_2| \widetilde{\mathcal{O}}^{\pm 1-h_1,1-\bar{h}_1}(w_1,\bar{w}_1)\mathcal{O}^{\mp, h_2,\bar{h}_2}(w_2,\bar{w}_2)|0_2\rangle. 
\end{align}
 Inserting~\eqref{Stwopoint}, and evaluating the integral using 
\begin{equation}
\frac{1}{2\pi i}\oint dz z^{h+m-1}(w-z)^{-2h} = \frac{w^{m-h}\Theta(-h-m)\Gamma(h-m)}{\Gamma(1-h-m)\Gamma(2h)}
\end{equation}
where $h+m\in\mathbb{Z}, h\in\mathbb{C}$, we find that the inner product is
\begin{eqnarray}\label{cmplexh}
\langle 0_2|\widetilde{\mathcal{O}}^{\prime \pm, 1-h_1,1-\bar{h}_1}_{n,\bar{n}}\mathcal{O}_{m,\bar{m}}^{\mp, h_2,\bar{h}_2}|0_2\rangle & = & \frac{\delta_{h_1-\bar{h}_1+h_2-\bar{h}_2}\mathcal{C}^\pm_{J_1}(\lambda_1)\delta(\lambda_1+\lambda_2)\Gamma(h_2-m)\Gamma(\bar{h}_2-\bar{m})}{\pi\Gamma(1-2h_2)\Gamma(2h_2)\Gamma(1-h_2-m)\Gamma(1-\bar{h}_2-\bar{m})}\cr
& \times & \Theta(n-h_2)\Theta(\bar{n}-\bar{h}_2)\delta_{m+n}\delta_{\bar{m}+\bar{n}}. 
\end{eqnarray}
Note that the above equation is only nonzero when $(h_2,\bar{h}_2) = (1-h_1,1-\bar{h}_1)$ and $(m,\bar{m})=(-n,-\bar{n})$. Therefore this formula identifies the adjoint of ${\mathcal{O}}^{\mp, h,\bar{h}}_{m,\bar{m}}|0_2\rangle$ as\footnote{The Kronecker delta functions here are valid precisely since $m+n\in (\mathbb{Z}-h)+(\mathbb{Z}+h) = \mathbb{Z}$. This would not have occurred if $n\in\mathbb{Z}-h$. For $h$ or $\bar h$ integers one may wish to absorb the $\Gamma$ function divergence into a renormalization of the operators.}
\be 
\langle 0_2|\widetilde{\mathcal{O}}^{\prime\pm, h,\bar{h}}_{-m,-\bar{m}}=({\mathcal{O}}^{ \mp, h,\bar{h}}_{m,\bar{m}}|0_2\rangle)^\dagger. 
\ee
One can also show that the two-point function can be reconstructed via a sum over all the modes, providing a check that our basis of descendants is complete.

This inner product brings us full circle and can be easily connected to the the bulk analysis and shadow product in the previous section. If we consider the primary states~\eqref{inket} and~\eqref{outbra}, we see that the inner product we define here is 
\be 
\langle 0_2|\widetilde{\mathcal{O}}^{ \prime\pm ,1-h_1,1-\bar{h}_1}_{1-h_1,1-\bar{h}_1}\mathcal{O}_{-h_2,-\bar{h}_2}^{\mp, h_2,\bar{h}_2}|0_2\rangle  =  \frac{\delta_{h_1-\bar{h}_1+h_2-\bar{h}_2}\mathcal{C}^\pm_{J_1}(\lambda_1)\delta(\lambda_1+\lambda_2)\Gamma(2\bar{h}_2)}{\pi\Gamma(1-2h_2)}. 
\end{equation}
Likewise we could have started with the bulk shadow product in~\eqref{eq:shadowsym}, multiplied it by the conformal factor $w_1^{2(1-h_1)}\bar{w}_1^{2(1-\bar{h}_1)}$, thereby taking the first field to the opposite pole, and then taken the limit $w_1,\bar{w}_1\rightarrow\infty$ and $w_2,\bar{w}_2\rightarrow 0$ to obtain the same expression. This should not be surprising since it follows directly from the CCFT state-operator correspondence. The same connection exists for descendants since considering arbitrary modes in~\eqref{cmplexh} will correspond to inserting $L_{-1}$ and $L_1$ in~\eqref{eq:shadowsym}, multiplying by the appropriate conformal factor and taking the limit.

Since the inner product here is  the standard BPZ product in CFT$_2$ it is manifest that 
$L_n^\dagger = L_{-n}$ and $\bar L_n^\dagger=\bar L_{-n}$, unlike the relation $L_n^\dagger=-\bar{L}_n$  encountered in the 4D bulk Klein-Gordon product. It is nevertheless instructive to derive the adjoint relations directly. As shown in Appendix~\ref{app:ln}
\begin{equation}\label{kid}
\langle 0_2|[L_{-n}\mathcal,\widetilde{\mathcal{O}}'^{\mp,h,\bar{h}}_{-m,-\bar{m}}]  =  -((h-1)n-m)\langle 0_2|\widetilde{\mathcal{O}}'^{\mp,h,\bar{h}}_{-m-n,-\bar{m}}.
\end{equation}
This allows us to compute $L_n^\dagger$ acting on the bra-state
\begin{equation}
    \bra{0_2} \widetilde{\mathcal{O}}'^{\pm,h, \bar{h} }_{m, \bar{m} } L_n^\dagger = \left( L_n \mathcal{O}_{m, \bar{m}}^{\mp,h, \bar{h}} |0_2 \rangle \right)^\dagger = ( (h-1) n - m) \left( \mathcal{O}^{\mp,h, \bar{h} }_{m + n, \bar{m} } \ket{0_2} \right)^\dagger = ( (h-1) n - m) \langle 0_2 | \widetilde{\mathcal{O}}'^{\pm,h, \bar{h}}_{-m-n, -\bar{m}}
\end{equation}
where in the second equality we used~\eqref{Lnmodes}. If we compare the above equation with~\eqref{kid}, we are able to conclude that
\begin{equation}\label{apromisekept}
L_n^\dagger = L_{-n}
\end{equation}
as promised.

This discussion could be traced backwards. Given that the BPZ construction implies the adjoint relation $L_n^\dagger = L_{-n}$, 
the inner products of descendants~\eqref{cmplexh} then follows, up to an overall normalization,  from the  
general SL$(2,\mathbb{C})$ relation~\eqref{Lnmodes}.

\section{Bulk versus boundary scattering}\label{sec:bulkvboundary}

Thus far we have discussed the inner products of quantum states in free field theory, as well as the novel relation between the bulk products on 4D states and the boundary inner product on 2D states. Here we make some observations on the associated relation between the bulk and boundary scattering problems. 

\begin{figure}[thb]
\begin{center}
\includegraphics[scale=0.5]{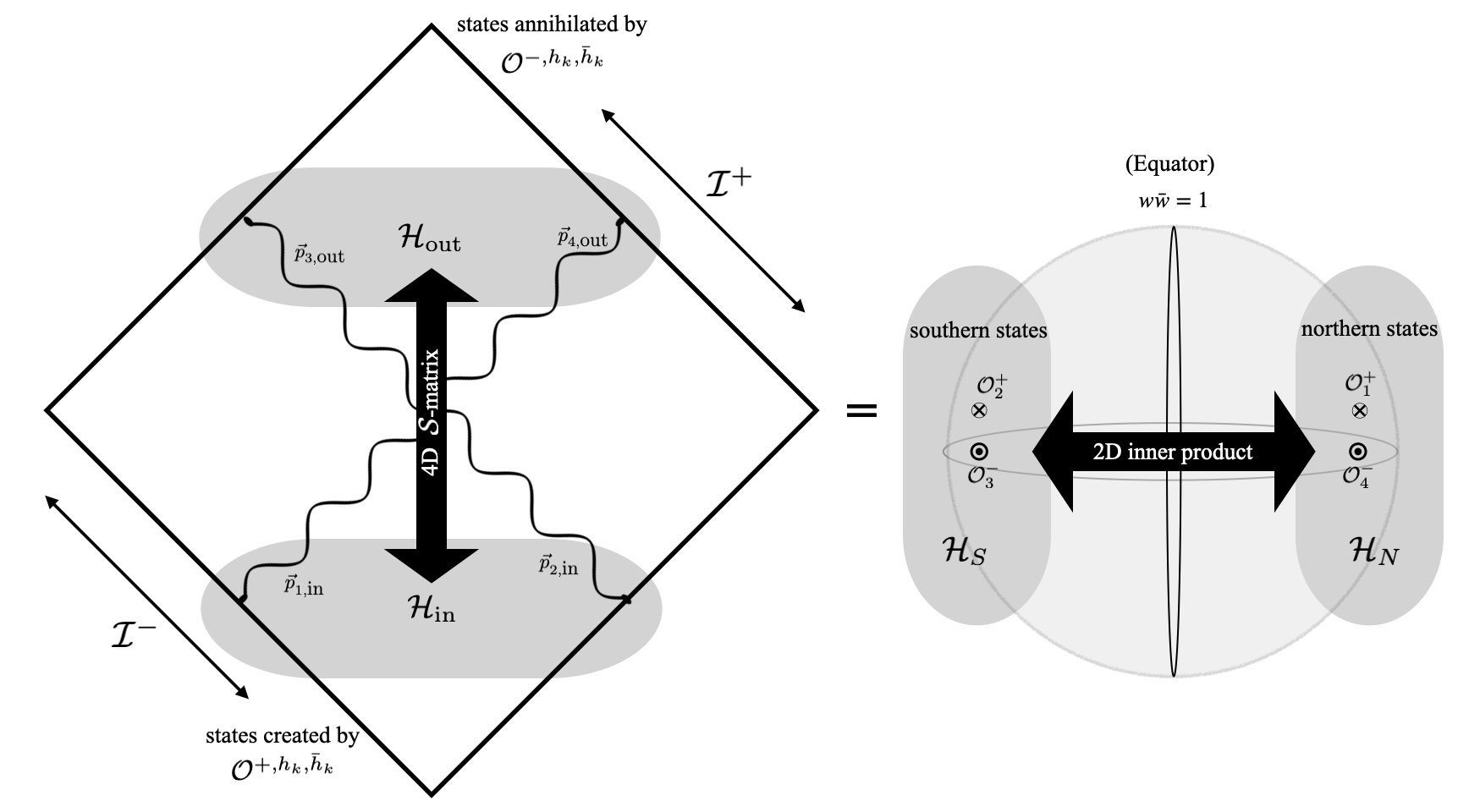}
\end{center}
\caption{On the left we depict the 4D scattering problem. Incoming states are in the Hilbert space $\mathcal{H}_{\rm{in}}$ at $\mathcal{I}^-$ and outgoing states are in the Hilbert space $\mathcal{H}_{\rm{out}}$ at $\mathcal{I}^+$. The bulk scattering problem is to relate these two Hilbert spaces. On the right we depict the 2D scattering problem. In the northern (southern) hemisphere we have northern (southern) states, those originating at the south (north pole) of $\mathcal{I}^-$, and they form the Hilbert space $\mathcal{H}_N$ ($\mathcal{H}_S$). The dots represent operators that annihilate the northern (southern) states on $\mathcal{I}^+$ and the crosses represent antipodally mapped operators that create the northern (southern) states on $\mathcal{I}^-$. The boundary scattering problem is to find the overlap between the states in these two Hilbert spaces.}
\label{hilbertpic}
\end{figure}

The bulk $N$-particle $\cal S$-matrix is a map $\ch_{\rm in}\to \ch_{\rm out}$ between a state in the incoming 4D Hilbert space on $\cal{I}^-$ and the outgoing Hilbert space on $\ci^+$ as shown in the left panel of Figure~\ref{hilbertpic}. In a conformal basis, the incoming $N$-particle states are created on $\cal{I}^-$ by products of positive frequency operators ${\mathcal{O}}^{ +, h_k,\bar{h}_k}(w_k, \bar w_k)$, $k=1,..N_{\rm in}$,  and are annihilated on $\ci^+$ by 
products of negative  frequency operators ${\mathcal{O}}^{ -, h_k,\bar{h}_k}(w_k, \bar w_k)$, $k=1,..N_{\rm out }$, with $N= N_{\rm in}+N_{\rm out}$.  The operators $\co^-$ ($\co^+$) are associated to $\ci^+$ ($\ci^-$). 

This same bulk $N$-particle $\cal S$-matrix can be expressed as a map between two 2D Hilbert spaces. To do so, we choose an equator on the celestial sphere. We take this to be  $w\bar w=1$ in our conventions and refer to $w \bar w<1$ as the northern hemisphere and $w \bar w >1 $ as the southern hemisphere as shown in the right panel of Figure~\ref{hilbertpic}. If there are no operator insertions on either hemisphere, starting from the vacuum at the north (south) pole, and evolving to the equatorial boundary of the northern (southern) hemisphere creates the vacuum state in the northern (southern) Hilbert space $\ch_{\rm N}$ ($\ch_{\rm S}$). If there are operator insertions at points on the hemispheres, evolution from the pole will create a corresponding state in the Hilbert space. Our inner product between states in either Hilbert space is then an $\cal S$-matrix element.

While the pairs $(\ch_{\rm in},\ch_{\rm out})$ and $(\ch_{\rm N},\ch_{\rm S})$ contain the same information, they clearly organize it quite differently. Whereas a single 4D Hilbert space contains either all the data on $\mathcal{I}^+$ or all the data on $\mathcal{I}^-$, a single 2D Hilbert space contains part of the data from $\mathcal{I}^-$ along with part of the data from $\mathcal{I}^+$. To see this, we should understand the operator insertions that appear on each 2D hemisphere. A single-particle 2D state in $\ch_{\rm N}$ can either be associated to an operator ${\mathcal{O}}^{ -, h_k,\bar{h}_k}(w_k, \bar w_k)$ or ${\mathcal{O}}^{ +, h_k,\bar{h}_k}(w_k, \bar w_k)$ with $w_k\bar w_k<1$. The operator ${\mathcal{O}}^{ -, h_k,\bar{h}_k}(w_k, \bar w_k)$ annihilates a bulk particle  on the northern hemisphere at $\mathcal{I}^+$ while ${\mathcal{O}}^{ +, h_k,\bar{h}_k}(w_k, \bar w_k)$ creates a north-moving bulk particle in the southern hemisphere at $\mathcal{I}^-$.

The bulk scattering problem can be phrased as: ``Given an incoming state in $\ch_{\rm in}$, what is the resulting outgoing state in $\ch_{\rm out}$?" The boundary scattering problem is ``Given a northern  state in $\ch_{\rm N}$, what is the resulting southern  state in $\ch_{\rm S}$?" The bulk interpretation of this second question is novel. Given the multi-particle state leaving the southern hemisphere of $\ci^-$, as well as the multi-particle state arriving in the northern hemisphere of $\ci^+$,  one must then reconstruct\footnote{In free field theory this would be an impossible task since they do not interact. But in gravity there are always phase shifts and other interactions between crossing states. This is reminiscent of the discussion of the black hole $\cal S$-matrix in~\cite{tHooft:1996rdg}.} the multi-particle state leaving the northern hemisphere of $\ci^-$, as well as the one arriving in the southern hemisphere of $\ci^+$. 

While a product of conformal primary operator insertions at distinct northern points $(w_k,\bar w_k)$ creates a state in $\ch_{\rm N}$ as one evolves from the north pole, it is often convenient in CFT$_2$ to consider instead arbitrary insertions of primary operators as well as their SL$(2,\mathbb{C})$ descendants\footnote{We could also consider Virasoro descendants, but the general Virasoro descendant of a single-particle primary is a multi-particle state including soft gravitons.}. These transform simply under the action of $(L_n, \bar L_n)$, and can be written\footnote{Of course in an interacting theory  the spectrum of such states will be corrected by the nontrival OPEs.} in the general form~\eqref{rdl} of operators acting on $|0_2\rangle$. Scattering amplitudes can then be identified with inner products of these ket-states with the analogous bra-states.

In conclusion the bulk to boundary dictionary in flat holography is more subtle than its familiar AdS counterpart. The 2D boundary states live on an oriented  circle in the celestial sphere and capture the information of both ingoing  and outgoing bulk  excitations which cross the celestial sphere inside this circle. The 2D boundary $\cal S$-matrix is a map from the 2D state on one side of the circle to the other, and corresponds to a bulk map determining south-to-north excitations from north-to-south ones. This is a significant reframing of the quantum gravity scattering problem in Minkowski space.

\section*{Acknowledgements}
This work was supported in part by DOE grant de-sc/0007870 to AS, NSF GRFP grant DGE1745303 to NM and the John Templeton and Gordon and Betty Moore Foundations via the Black Hole Initiative. The authors have benefited from enlightening discussions with Alex Atanasov, Adam Ball, Arindam Bhattacharya, Scott Collier, Alfredo Guevara, Mina Himwich, Walker Melton, Colin Nancarrow, Rajamani Narayanan, Aditya Parikh, Monica Pate, Ana Raclariu, Tomasz Taylor, Neeraj Tata, and Xi Yin.

\appendix

\section{Conformal generators on adjoint modes}\label{app:ln}
In this section we will show how, for $n = 0, \pm 1$, the action of $L_n$ and $L_n^\dagger$ on the modes $\mathcal{O}^{\mp,h, \bar{h}}_{m, \bar m}$ and $\widetilde{\mathcal{O}}'^{\pm,h, \bar{h}}_{-m, -\bar m}$ can be derived straight from the two-point function. This might seem like a peculiar exercise. However, we do this in order to establish that these equations apply in the CCFT case, where we \textit{start} with the two-point function and then infer the properties of the Hilbert space from it.

Defining
\begin{eqnarray}
\langle \widetilde{\mathcal{O}}'^{\pm}_1\mathcal{O}_2^{\mp}\rangle & \equiv & \langle 0_2| \widetilde{\mathcal{O}}'^{\pm,1-h_1,1-\bar{h}_1}(w_1,\bar{w}_1)\mathcal{O}^{\mp,h_2,\bar{h}_2}(w_2,\bar{w}_2)|0_2\rangle\cr
& = & \frac{\Gamma(2\bar{h}_2)\mathcal{C}^{\pm}_{J_1}(\lambda_1)\delta(\lambda_1+\lambda_2)}{\pi\Gamma(1-2h_2)}\frac{w_1^{2h_2}\bar{w}_1^{2\bar{h}_2}\delta_{h_1+h_2-\bar{h}_1-\bar{h}_2}}{(w_1 - w_2)^{2h_2} (\bar{w}_1 - \bar{w}_2 )^{2 \bar{h}_2 }}
\end{eqnarray}
via explicit computation one can show that
\begin{equation}\label{Lnoutside}
(h_2(n+1)w_2^n+w_2^{n+1}\partial_{w_2})\langle \widetilde{\mathcal{O}}'^{\pm}_1\mathcal{O}^{\mp}_2\rangle = -((1-h_1)(n-1)w_1^n+w_1^{n+1}\partial_{w_1})\langle \widetilde{\mathcal{O}}'^\pm_1\mathcal{O}_2^\mp\rangle
\end{equation}
holds for $n=0,\pm 1$. This can be thought of as performing the computation of $\langle \widetilde{\mathcal{O}}'^\pm_1 L_n \mathcal{O}_2^\mp\rangle$ in two different ways, by acting $L_n$ on the right and on the left. This implies that
\begin{eqnarray}\label{Ln_in}
[L_n, \widetilde{\mathcal{O}}'^{\pm,1-h,1-\bar{h}}(z,\bar{z})] & = & (1-h)(n-1)z^n\widetilde{\mathcal{O}}'^{\pm,1-h,1-\bar{h}}(z,\bar{z})+z^{n+1}\partial_z\widetilde{\mathcal{O}}'^{\pm,1-h,1-\bar{h}}(z,\bar{z})
\end{eqnarray}
if one also assumes that for $n = 0, \pm1$, $L_n$ annihilates the vacuum $\langle 0_2 | L_n = 0$. In a more standard operator derivation, the difference between the commutator of $L_n$ with $\mathcal{O}^{\pm,h,\bar{h}}(w,\bar{w})$ and $\mathcal{O}'^{\pm,h,\bar{h}}(z,\bar{z})$ is due to the definition of the out states, as
\begin{equation}
    [L_n,  \mathcal{O}'^{\pm,h,\bar{h}}(z, \bar{z}) ] = z^{2h}\bar{z}^{2\bar{h}}[L_n, \mathcal{O}^{\pm,h,\bar{h}}(z, \bar{z})].
\end{equation}
Using the mode expansions gives the following relation for the modes
\begin{eqnarray}\label{eq_Lonmode}
     \langle 0_2|[L_n\mathcal,\widetilde{\mathcal{O}}'^{\pm,h,\bar{h}}_{-m,-\bar{m}}] & = & ((h-1)n+m)\langle 0_2|\widetilde{\mathcal{O}}'^{\pm,h,\bar{h}}_{-m+n,-\bar{m}}
\end{eqnarray}
for $n = 0, \pm 1$. The above equation is used in the derivation of ~\eqref{apromisekept}, the adjoint equation $L_n^\dagger = L_{-n}$. In a standard unitary CFT, $h^\dagger = h$ because $h$ is real. However, as we have constructed an inner product which is not sesquilinear (where the dual state is complex conjugated) but instead bilinear, then the adjoint of a scalar is itself. In that case, $h^\dagger = h$ will be true even if $h$ is complex. This is indeed the case in our CCFT inner product, which is bilinear in the fields.

\bibliography{cpb}
\bibliographystyle{utphys}

\end{document}